\documentclass{aa}
\usepackage{psfig}
\usepackage{graphicx}

\def\bron{GRS 1747-312}
\def\ecs{erg~cm$^{-2}$s$^{-1}$}
\def\lum{erg~s$^{-1}$}

\begin{document}

\title{A peculiar type-I X-ray burst from \bron}

\titlerunning{Peculiar X-ray burst from \bron} 
\authorrunning{J.J.M. in 't Zand, T.E. Strohmayer, C.B. Markwardt \& J. Swank}

\author{
J.J.M.~in~'t~Zand\inst{1,2}
\and T.E.~Strohmayer\inst{3}
\and C.B. Markwardt\inst{3,4}
\and J. Swank\inst{3}
}

\offprints{J.J.M. in 't Zand, email {\tt jeanz@sron.nl}}

\institute{     SRON National Institute for Space Research, Sorbonnelaan 2,
                NL - 3584 CA Utrecht, the Netherlands 
	 \and
                Astronomical Institute, Utrecht University, P.O. Box 80000,
                NL - 3508 TA Utrecht, the Netherlands
	 \and
                NASA Goddard Space Flight Center, Code 662, Greenbelt,
                MD 20771, U.S.A.
         \and
                Dept. of Astronomy, University of Maryland, College Park,
                MD 20742, U.S.A.
	}

\date{Received, accepted }

\abstract{We report the serendipitous detection with the Rossi X-ray
Timing Explorer of a long and peculiar X-ray burst whose localization
is consistent with one known X-ray burster (\bron) and which occurred
when that source was otherwise quiescent. The peculiar feature
concerns a strong radius expansion of the neutron star photosphere,
which occurred {\em not} within a few seconds from the start of the
burst, as is standard in radius-expansion bursts, but 20~s later.
This suggests that two different layers of the neutron star may have
undergone thermonuclear runaways: a hydrogen-rich and a hydrogen-poor
layer. The reason for the delay may be related to the source being
otherwise quiescent.  \keywords{accretion, accretion disks -- globular
clusters: individual: Terzan~6 -- X-rays: binaries -- X-rays: bursts
-- X-rays: individual: \bron\ }}

\maketitle 

\section{Introduction}
\label{intro}

\bron\ is a transient X-ray source that is located in the core of the
9.5~kpc distant globular cluster Terzan~6 (Predehl et al. 1991;
Pavlinsky et al. 1994; Verbunt et al. 1995; In 't Zand et
al. 2003). The heart of the source is a neutron star in a binary
system with a low-mass, Roche lobe filling star. X-ray outbursts occur
every 4.5 months and last 1 month; the neutron star is completely
eclipsed by the companion star every 12.4~hrs (In 't Zand et al. 2000,
2003). The neutron star nature was established through the detection
of seven type-I X-ray bursts. Such bursts are clear markers of
thermonuclear runaway processes on the surfaces of neutron stars (for
reviews, see Lewin et al. 1993 and Strohmayer \& Bildsten 2003).  All
bursts were short-lived and occurred while the source was otherwise
X-ray active. In April 2002 \bron\ happened to be in the field of view
of the Proportional Counter Array (PCA) on the Rossi X-ray Timing
Explorer (RXTE) while it was pointed at another target, and a long
type-I X-ray burst was detected with peculiar features that may add to
the understanding of unstable nuclear burning on neutron star
surfaces. Here we discuss this event.

\section{Observations}
\label{pca}

\begin{figure}[t]
\psfig{figure=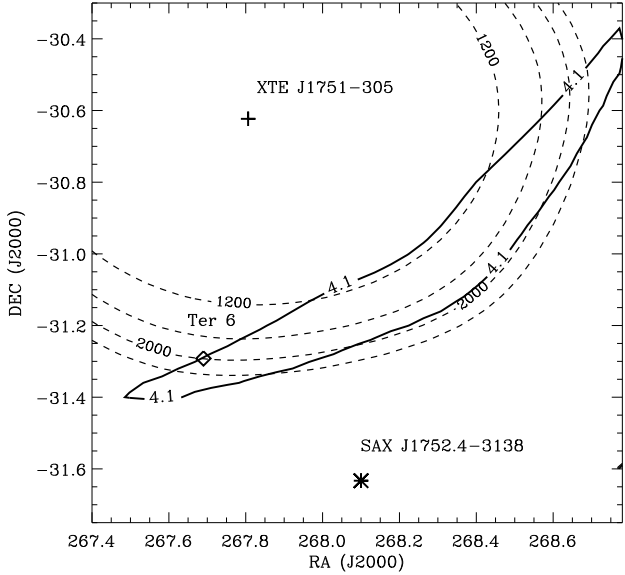,width=\columnwidth,clip=t}
\caption{Map of the burst detected on April 30.9, 2002. The thick
contour borders the region from where the burst originated with 90\%
confidence. The cross indicates the position of XTE~J1751-305, the
target of this observation, and the diamond the position of
Terzan~6. The burst does not originate from XTE~J1751-305, while
\bron\ in Terzan 6 is plausible.  The dashed contours indicate what
the background-subtracted peak photon count rate per PCU would have been
if the true location were on that contour.
\label{figburstpos}}
\end{figure}

The PCA (for a detailed description, see Jahoda et al. 1996) on RXTE
comprises 5 co-aligned Proportional Counter Units (PCUs) that are
sensitive to 2 to 60 keV photons. The total collecting area is
6500~cm$^2$. The spectral resolution is 18\% full-width at half
maximum (FWHM) at 6 keV and the field of view is 1$^{\rm o}$
FWHM. During April 4 through 30, 2002, the PCA was employed in a
target-of-opportunity (ToO) program on the newly discovered accreting
millisecond pulsar and transient source XTE~J1751-305 (Markwardt et
al. 2002) which is located 40\arcmin\ from \bron. The program
accumulated 398~ks of net exposure time.  \bron\ was not in outburst
during these serendipitous observations. On the very last day of
observations an X-ray burst was detected.  Figure~\ref{figburstpos}
shows the result of the localization of this burst following a method
(Strohmayer et al. 1997) that employs the small misalignments of the
three PCUs that were active during the burst (PCUs 0, 2 and 3). While
XTE~J1751-305 is excluded as the origin of the burst at the $>90$\%
confidence level, the error region is consistent with \bron\ being the
origin. Furthermore, no other X-ray burster is known within the 90\%
confidence contour. It should be mentioned that the light curve of
XTE~J1751-305 (see Fig.~1 in Markwardt et al. 2002) shows a small
outburst to a level of 0.3~mCrab starting three days before the
burst. Given the collimator response to \bron\ in these observations,
this translates to a level of 1.1~mCrab if the outburst would have
been due to \bron. Since this level is excluded by independent
observations of \bron\ (see Fig.~\ref{figbulgelc} in which 1.1~mCrab
is equivalent to 12 phot~s$^{-1}$ per 5 PCUs), \bron\ is ruled out as the
source of the outburst.

\begin{figure}[t]
\psfig{figure=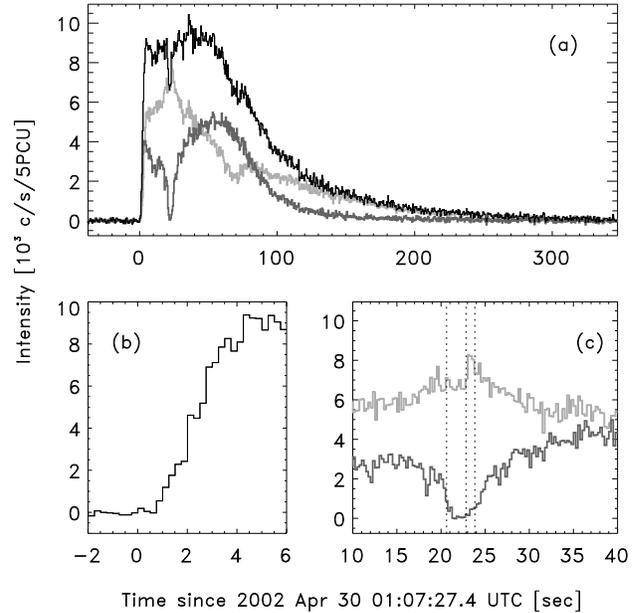,width=1.0\columnwidth,clip=t}
\caption{(a) Lightcurves of $<10$~keV photons (light grey curve),
$>$10~keV photons (dark grey curve) and all photons (black curve), at
0.5~s resolution; (b) lightcurve of burst onset for full bandpass at
0.25~s resolution; (c) $<10$~keV (light grey) and $>10$~keV (dark
grey) lightcurve, zoomed in on softening episode at 0.25~s resolution;
the vertical dashed lines delimit time intervals for which spectra are
shown in Fig.~\ref{fig3sspectrum}. In all panels, the background has
been subtracted, all active PCUs were used (0, 2 and 3), and a
correction has been applied for the collimator response (0.272 for
PCUs 0 and 2 and 0.298 for PCU3). The orbital phase with respect to
mid-eclipse is 0.48 (i.e., the companion star is on the far side of
the neutron star).
\label{fig5lc}}
\end{figure}

\section{Time-resolved spectroscopy and timing analysis}

Figure~\ref{fig5lc} shows the light curve of the burst in various
bandpasses. It is characterized by a duration of 320$\pm10$~s and a
rise time of 3~s. The full-bandpass light curve shows a plateau for
the first $\approx$70~s strongly suggesting that the Eddington limit
is reached. There is one particularly peculiar feature about this
burst: from 18 to 27~s after burst onset, the $>$10~keV intensity
shows a sharp drop, indicating a strong softening of the spectrum. 

\begin{figure}[t]
\psfig{figure=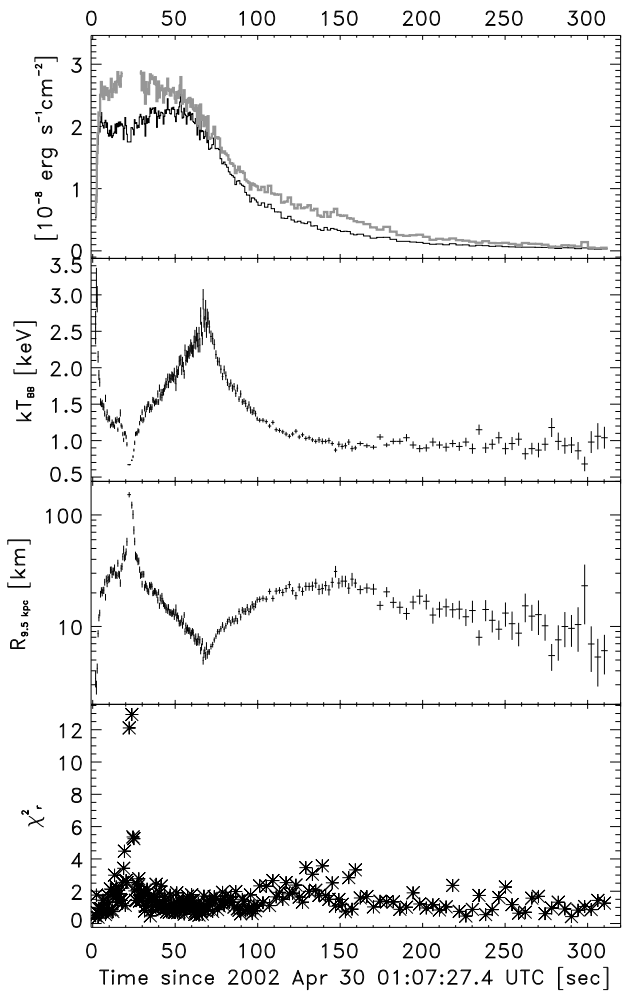,width=1.0\columnwidth,clip=t}
\caption{Time-resolved spectroscopy of the burst, modeling with an
absorbed blackbody spectrum, leaving free a single value for $N_{\rm
H}$ over all intervals. The top panel shows the predicted unabsorbed
bolometric flux (grey curve) and the same flux within the instrument's
bandpass (3--20 keV; solid curve). The fitted overall value for
$N_{\rm H}$ is 6$\times10^{22}$~cm$^{-2}$.
\label{fig5sp1}}
\end{figure}

\begin{figure}[t]
\psfig{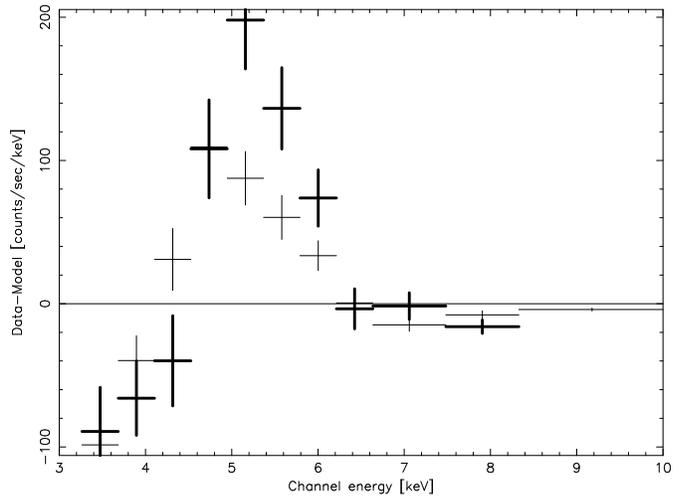}
\caption{Residuals with respect to a black body fit of spectra taken
during time intervals specified in Fig.~\ref{fig5lc}c (thin-drawn
symbols are for first interval and fat for second).
\label{fig3sspectrum}}
\end{figure}

We generated time-resolved spectra for the burst, from event mode data
that have a resolution of 2$^{-13}$~sec, 64 energy channels, and no
layer resolution. The time resolution of the spectra was chosen to
vary over the burst, in order to keep the statistical quality of the
spectra at an approximately constant level. The time resolution varied
from 0.5 to 4.0~s. A background spectrum was determined from data
preceding the burst from $-40$ to $-440$~s, and a correction was
applied for the deadtime. The spectra were, between 3 and 20~keV,
fitted with an absorbed single-temperature black body model with
varying temperature and normalization. The absorption was modeled
following Morrison \& McCammon (1983), leaving free one value for the
hydrogen column density $N_{\rm H}$ for all spectra.  Furthermore,
spectral channels were combined so that each spectral bin contained at
least 15 photons, to ensure Gaussian statistics and the applicability
of the $\chi^2$ statistic. The fitted model is satisfactory for all
time intervals, except during the brief episode of strong
softening. The fit results are presented in Fig.~\ref{fig5sp1}.  The
residuals of the data with respect to the model are shown in
Fig.~\ref{fig3sspectrum} for the two time intervals specified in
Fig.~\ref{fig5lc}. In these intervals there is a strong hint of a
feature peaking at 5~keV. The residual may be fitted by a broad
Gaussian function ($\chi^2_\nu$ reduces from 11.6 [22 dof] to 1.9 [19
dof]). The model fits to a centroid of 4.9$\pm0.1$~keV, a FWHM of
1.2$\pm0.2$~keV, and an intensity of
0.18$\pm0.03$~phot~s$^{-1}$cm$^{-2}$. At its peak, approximately 30\%
of the bolometric flux is predicted to be due to this component.  The
sudden increase of the $<10~$keV photon rate at 23~s (see panel (c) in
Fig.~\ref{fig5lc}) can be explained by a sudden temperature rise
from $0.57\pm0.03$ to $0.71\pm0.02$~keV.  Although the Gaussian fit is
best, considerable improvement may also be achieved by allowing
$N_{\rm H}$ to vary; $\chi^2_\nu$ reduces to 2.9 (21 dof) and $N_{\rm
H}$ goes to $(4-5)\times10^{23}$~cm$^{-2}$. However, the inferred
radius then becomes excessively large at 1.7$\times10^3$~km at
9.5~kpc.

The fit results show a strongly varying blackbody radius, with a
peak-to-peak variation of at least a factor of 40, that is
anti-correlated with the temperature variation. During the first
$\approx$70~s this is accompanied by a fairly constant luminosity.
This is a clear detection of photospheric radius expansion (PRE) due
to near-Eddington luminosities. The radius increase beyond 70~s goes
with a decay in luminosity. This is often seen in PRE bursts and is
assumed to be related to non-Planckian spectra (e.g., Kaptein et al.
2000, van Straaten et al. 2001). It is not thought to be related to
true radius expansion. At the end of the burst, the radius decreases
to a similar level as in the bursts that have been detected with the
PCA when \bron\ was active (In~'t~Zand et al. 2003).

During the brief episode of strong softening, the expansion is strongest.
Given the fact that this appears to be an isolated event, the
suggestion is strong that in fact we are dealing with two different
phases of radius expansion: the slow 70~s lasting expansion with a
factor of a few, and the fast 8~s lasting expansion with a factor of
a few tens.

During the fast radius expansion, the temperature drops significantly
below the low-energy threshold of the instrument and a significant
fraction of blackbody radiation escapes detection. Consequently,
measurements of the radius and bolometric flux become rather
inaccurate, particularly if one would chose to leave free $N_{\rm H}$
(which was {\em not} done in the analysis shown in
Fig.~\ref{fig5sp1}).  The bolometric flux saturates during the first
70~s of the burst at a level of about 2.5$\times10^{-8}$~\ecs\
(bolometric), disregarding the fast radius expansion phase during
which bolometric corrections are too inaccurate.

We note that no burst oscillations were detected during this burst.
The upper limit to the amplitude is 16\% on a timescale of 4~s
(combining the data from PCUs 0, 2 and 3; for details on the method,
see In~'t~Zand et al. 2003).

The burst lasted at least ten times longer ($320\pm10$~s to reach the
background level) than the other four bursts that were detected from
\bron\ with the PCA (durations between 10 and 32~s; In 't Zand et
al. 2003). However, the peak flux is, within 20\%, identical to the
brightest PRE burst of the other four. This supports the association
of the burst with \bron, as does the fact that its longevity is
expected for low accretion rates from hydrogen-rich donors (e.g.,
Fujimoto et al. 1981). \bron\ was not in outburst during the
burst. The prior outburst was last detected above 1~mCrab 26 days
earlier and the following outburst started 28 days afterwards (see
Fig.~\ref{figbulgelc}).  The accretion rate is suggested to have been
a factor of at least 40 smaller than during the peak of the
outbursts. Similar behavior was observed in other bursters as well,
for instance SAX~J1808.4-3658 (In~'t~Zand et al. 2001) and 2S~1711-339
(Cornelisse et al. 2002).

\section{Discussion}
\label{discuss}

The burst has a relatively strong photospheric radius expansion.  Such
bursts, with forty-fold or more expansion, are rather rare. We are
aware of four other bursters that exhibit such strong radius expansion
and, remarkably, three are located in globular clusters: 2S~1715-321
(Hoffman et al. 1978; Tawara et al. 1984), 4U~2129+11 in M15 (Van
Paradijs et al. 1990; Smale 2001), 4U~1724-307 in Terzan~2 (Molkov et
al. 2000) and 4U~1820-303 in NGC~6624 (Strohmayer \& Brown 2002). The
bursts from 4U~1724-307 occur much more frequently than in the other
sources, about once every 4 days (Kuulkers et al. 2003). Additionally,
one burst with a strong expansion was observed from an ill-determined
position and identification with a particular source was impossible
(Hoffman et al. 1978; Lewin et al. 1984). The expansions seen with the
PCA in 4U~1724-307 and 4U~1820-303 were so strong that no radiation
remained in the bandpass, not even the persistent emission otherwise
present for these sources.

What sets the burst from \bron\ apart from the other aforementioned
bursts is that the strong expansion phase did {\em not} occur at the
start of the burst, but is delayed by 20~s. It lasted less than 10~s
while (Eddington-limited) fluxes appear to remain at the same level
within 30\% from 20~s before to 35~s after the strong radius
expansion. Thus, the suggestion is strong that there must have been an
extra radiation pulse for $\approx$10~s whose energy was immediately
transformed to kinetic energy of a quickly expanding photosphere. We
propose that this extra energy is due to a thermonuclear flash which
occurred in a separate and deeper layer than the one where the long
flash originated. The duration suggests that this separate layer
consisted of hydrogen-poor material.  The layer may have formed either
through stable hydrogen burning during the previous outburst or
(invisible) unstable hydrogen burning after that (Fujimoto et
al. 1981). In the latter case, low-level accretion must have been
going on after the outburst.

\begin{figure}[t]
\psfig{figure=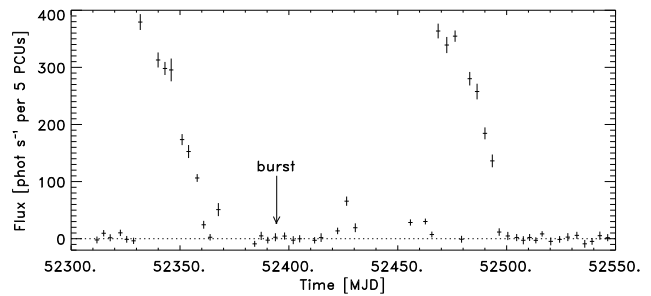,width=\columnwidth,clip=t}
\caption{Since February 1999 the Proportional Counter Array scans a
$16\degr\times16\degr$ field centered on the Galactic center,
encompassing \bron, twice a week (Swank \& Markwardt 2001; Markwardt
\& Swank 2002). Fluxes of the point sources in this field are
monitored, most for 60~s per scan. This figure shows the light curve
for \bron, zoomed in at around the time of the burst (between 26
January 2001 and to 3 October 2002). These observations did not cover
the burst. Gaps in this plot are caused by sun passages. The $3\sigma$
upper limit to the persistent emission at the time of the burst, if
extrapolated to 0.1-200~keV through the active-state spectrum, is
$6\times10^{35}$~\lum.
\label{figbulgelc}}
\end{figure}

What physical process or circumstance is responsible for the
unprecedented delay? A strong hint is provided by one characteristic
which sets \bron\ apart from the others: \bron\ is in quiescence,
while the other sources are persistently bright.  The difference in
luminosity is at least a factor of 10 (ranging from an upper limit of
$6\times10^{35}$~\lum\ in \bron, see Fig.~\ref{figbulgelc}, through
$7\times10^{36}$~\lum\ for 4U~1724-307 [Molkov et al. 2000] and
4U~2129+11 [Van Paradijs et al. 1990] to $2\times10^{37}$~\lum\ for
4U~1820-303 [Strohmayer \& Brown 2002]).  The lower accretion rate in
\bron\ implies that the temperature in the upper neutron star layers
is lower. This may delay the ignition conditions for the helium layer.
Perhaps this is related to a relatively slow onset of the second
convection stage that is predicted to occur in these types of X-ray
bursts following the growth of 3$\alpha$ reactions in the accreted
layer (Hanawa \& Fujimoto 1986).

The results on \bron\ put the other bursts with strong expansion
phases into a new perspective.  The strong radius expansion phases
that were detected in 4U~2129+11 (Van Paradijs et al. 1990) and
4U~1724-307 (Molkov et al. 2000) are accompanied by a prolonged but
smaller expansion afterwards.  Thus, there is the suggestion that also
in these bursts two separate layers may be ignited. This was perhaps
not recognized previously in the literature, because in those cases
the two layers ignited near-to simultaneously.

We note that long flashes (i.e., longer than 2~min) without strong
radius expansion have been observed in several low-luminosity systems
(e.g., Kaptein et al. 2000, In~'t~Zand et al. 2002). Perhaps in these
systems the accretion rate never becomes high enough to induce stable
hydrogen burning in the upper layers. Thus, a pure helium layer is
less likely to be present in these systems, precluding the potential
of a strong radius expansion.

The burst from \bron\ is fundamentally different from another peculiar
burst from 4U~1636-536 which was observed in August 1985 (Van Paradijs
et al. 1986). This 1-min long burst was triple peaked with the third
peak occurring 13~s after the first peak and the peak fluxes of similar
magnitude but at least a factor of four below the Eddington limit. No
photospheric radius expansion was detected. Van Paradijs et
al. speculate that this behavior is due to recurrent energy release in
quick succession at different locations on the surface of the neutron
star.

Another interesting detail about this burst is that it apparently
exhibited broad line emission. The nature of this feature is unclear;
the centroid energy of 4.8~keV is not coincident with that of any
known transition of an abundant element, Also, it requires
unexplainable redshifts from the Fe-K complex, particularly since the
photosphere radius is large when this feature was present, and we are
apprehensive to attribute it to Fe-K emission although such emission
was occasionally seen in other bursts (e.g., in the superburst of
4U~1820-303, see Strohmayer \& Brown 2002). Interestingly, this is not
the first time that such a broad emission feature has been seen in a
burst at this energy. The analysis by Van Paradijs et al. (1990) of
the burst from 4U~2129+11 in M15 showed severe expansion of at least a
factor of 80 and during the strongest phase of the expansion, a broad
emission feature was apparent that extended from 4 to 8 keV. They
exclude explanations that involve a broad emission line, reflection
and Comptonization.

\acknowledgement We thank Ed Brown and Walter Lewin for useful discussions.

\end{document}